\documentclass[11pt,epsfig]{article}
\usepackage{epsfig}
\newcommand{\ttilt}{{t_{\rm tilt}}}
\newcommand{\jdegr}{${}^\circ$}
\title{{\bf Tidal Torques and Galactic Warps}}
\author{J.~Bailin, M.~Steinmetz\\
\vspace{0.1cm}\\
\normalsize Steward Observatory, 933 N Cherry Ave, Tucson AZ, USA\\
}
\date{}
\begin{document}
\maketitle
\def\bull{\vrule height .9ex width .8ex depth -.1ex}
\makeatletter
\def\ps@plain{\let\@mkboth\gobbletwo
\def\@oddhead{}\def\@oddfoot{\hfil\tiny
``Dwarf Galaxies and their Environment'';
International Conference in Bad Honnef, Germany, 23-27 January 2001}%
\def\@evenhead{}\let\@evenfoot\@oddfoot}
\makeatother


\begin{abstract}\noindent

We investigate the tilting and warping of galactic disks in response to tidal
torquing.  The strength of the torque is determined from cosmological N-body
simulations. We find the tidal torques to be dominated by substructure in the
galactic halo, such as dwarf satellites, and by a misalignment between the disk
angular momentum and the figure axes of the dark matter halo. The radial
dependence of the torque can  approximated by a power law of index -2.5.
A massless disk subjected to such a torque develops a trailing warp similar to those
seen in a large number of disk galaxies, i.e.~the inner regions of
the disk tilt faster than the outer disk. In the case of massive disks, self
gravity causes the inner disk to stay locally flat and only the outer regions
show the signatures of warps. The radius outside of which a massive disk is warped
depends on the local surface density of the disk and on 
the product of the strength of the torque and
the exposure time to this torque.
\end{abstract}

\section{Introduction}
Many edge-on disk galaxies show integral-sign warps, where the majority of the disk
is planar but where the outer region of the disk lies above the plane on one side
of the galaxy and below the plane on the other (e.g.~Binney~1992). Most extended HI disks
appear warped (e.g.~Briggs~1990) and half of all disk galaxies have
optical warps (Reshetnikov~\& Combes~1998).

Various methods have been proposed for creating and maintaining warps, such as
normal bending modes (e.g.~Sparke~\& Casertano~1988) and disks askew in flattened
dark matter halos (Toomre~1983; Dekel~\& Shlosman~1983). Motivated by
the idea that infalling material will 
alter the direction of the angular momentum of a galaxy (Quinn~\&
Binney~1992), Ostriker~\& Binney~(1989) studied the reaction of massive rings to
a slewing disk potential. They found that warps occurred in regions of low
surface density.  Also motivated by the cosmic infall of angular momentum,
Debattista~\& Sellwood~(1999) found that when the angular momenta of a halo and
disk are misaligned, dynamical friction between them can produce a warp.

In a cosmological setting, a galactic disk is expected to continuously
experience tidal torques from a variety of sources, the three more important
being the distribution of mass in the local environment of a galaxy,
substructure in the dark matter halo such as dwarf satellites and high velocity
clouds (HVCs), and a misalignment between the disk angular momentum and the
figure axes of the dark matter halo. In this paper, we use cosmological N-body
simulations to deduce what gravitational tidal torques a typical galaxy
experiences from these three sources, and study whether these torques provide a
possible origin for warped disks.

\section{Torques in a Cosmological Context}
\label{torque section}

\begin{figure*}[tbp]

\vskip-3cm\hskip-5cm\psfig{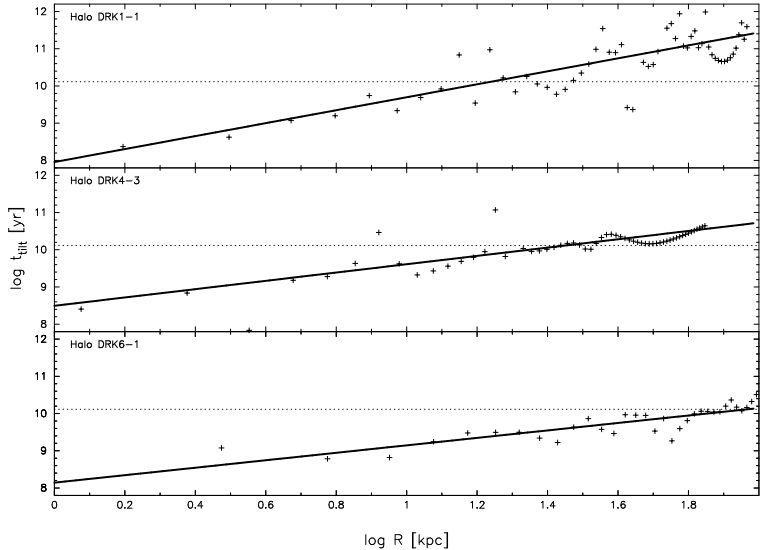}

\caption[]{Gravitational torques, expressed in terms of $\ttilt$,
at 60 evenly-spaced radii for three ~typical halos.
The solid lines are power law fits for the data points at
\protect\( R\geq 1.5\: \mathrm{kpc}\protect \).
Dotted lines correspond to the Hubble time (13~Gyr).\label{profiles.ps}}
\end{figure*}

We determine the strength and radial dependence of the torque by analyzing a
sample of 19 halos formed in an N-body simulation of structure formation in
the standard cold dark matter scenario ($\Omega_0=1$; for details, see Navarro~\&
Steinmetz~1997). Each halo consists of at least 1000 particles and has been
selected not to be significantly disturbed by recent merger events. We assume a
disk to form in the plane perpendicular to the angular momentum of the dark
matter halo.

The magnitude and radial dependence of the torque from all sources is determined
by calculating the gravitational force component that is exerted by all
particles in the simulation and that is acting perpendicular to the disk. The
torques are expressed in terms of the tilting timescale $\ttilt(r)$ defined as
the time required for a solid massless ring of radius $r$ to react to the torque
by tilting by one~radian.

We separate the mass in the simulation into that inside the halo (defined by the
virial radius $r_{200}$) and that outside the halo and compare the torques due to
each. We find the torque due to the mass outside the halo to be negligible,
indicating that galaxies in the local universe and cosmic tidal shear do not
contribute to the torque a (field) disk galaxy  experiences.

The torques inside three sample halos are shown in Figure~\ref{profiles.ps}.
The solid line is a power-law fit for each halo. The fits reproduce the broad
behavior of the torquing forces, but there is 
significant deviation at specific radii.
This power-law type radial dependence is characteristic of misalignments 
typical for dark matter halos that form by hierarchical clustering
(Cole~\& Lacey~1996; Warren et al.~1992), with mild misalignments of typically
20\jdegr\ for a moderately flattened ($b/a\approx0.8$) dark matter halo.
The deviations in the radial behavior from the power-law type dependence are due
to  substructure within the halo, such as satellite dwarfs, and, to some extent, due to numerical
noise.

The inner regions of the halo exhibit stronger torques 
(or equivalently shorter tilting timescales)
than the outer parts, i.e., the inner disk should tilt faster.
This is because the torquing force, which scales roughly as the density,
increases more rapidly toward the centre than the disk's
ability to resist the torque due to its angular momentum.
Also note that the timescales
are less than a Hubble time (indicated by the dotted lines)
over much of the disk, so we should see the effects
of these torques in a large number of observed galaxies, consistent with
observational findings (Briggs~1990, Reshetnikov~\& Combes~1998).

\section{Reaction of the Disk}

We study the reaction of a disk to these torques by performing numerical N-body
simulations of a massive Milky Way type galactic disk (maximum circular
velocity $v_{\rm max}=233\,$km/s,
scale length $r_d=3.5\,$kpc, vertical scale height $h_z=325\,$pc)
 subject to the torques found in section~\ref{torque
section}. Equilibrium disk models of disk mass $1\times 10^{10}\> M_\odot$,
$3\times 10^{10}\> M_\odot$, and $5.6\times 10^{10}\> M_\odot$
in a
static spherically-symmetric NFW halo potential (Navarro, Frenk \& White 1997)
with concentration parameter $c=15$ and 
a virial velocity of $v_{200}=175\,$km/s  are
constructed using the method of Hernquist (1993). Each disk contains 16384
particles.
The models were evolved using GRAPESPH (Steinmetz 1996) for 2~Gyr, which
took 5000--7000 timesteps depending on the model. A plummer
softening of 0.3~kpc has been used.

\begin{figure*}[tbp]
\centering{
\vbox{\psfig{figure=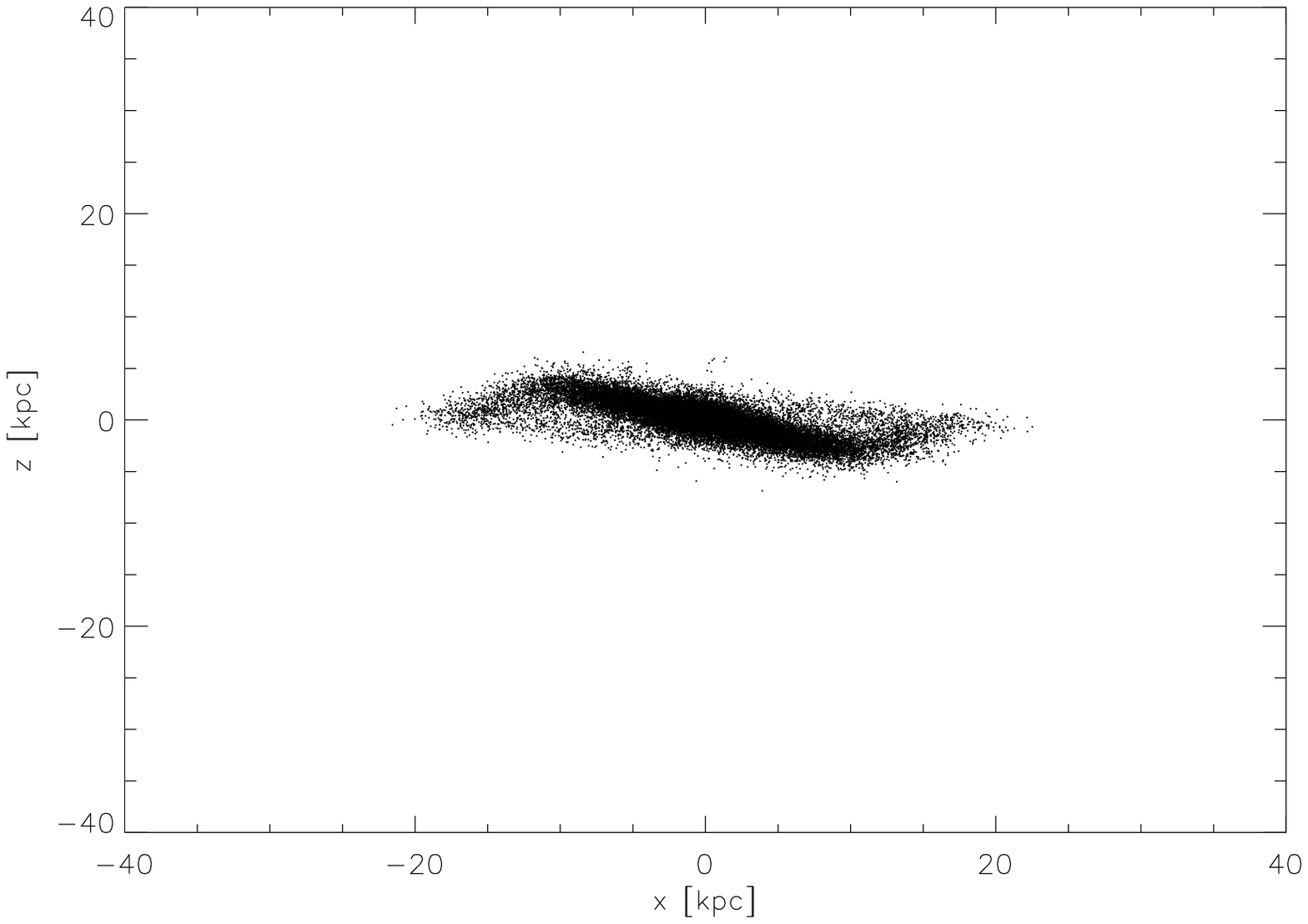,width=8.0cm,angle=0.0,clip=}
\psfig{figure=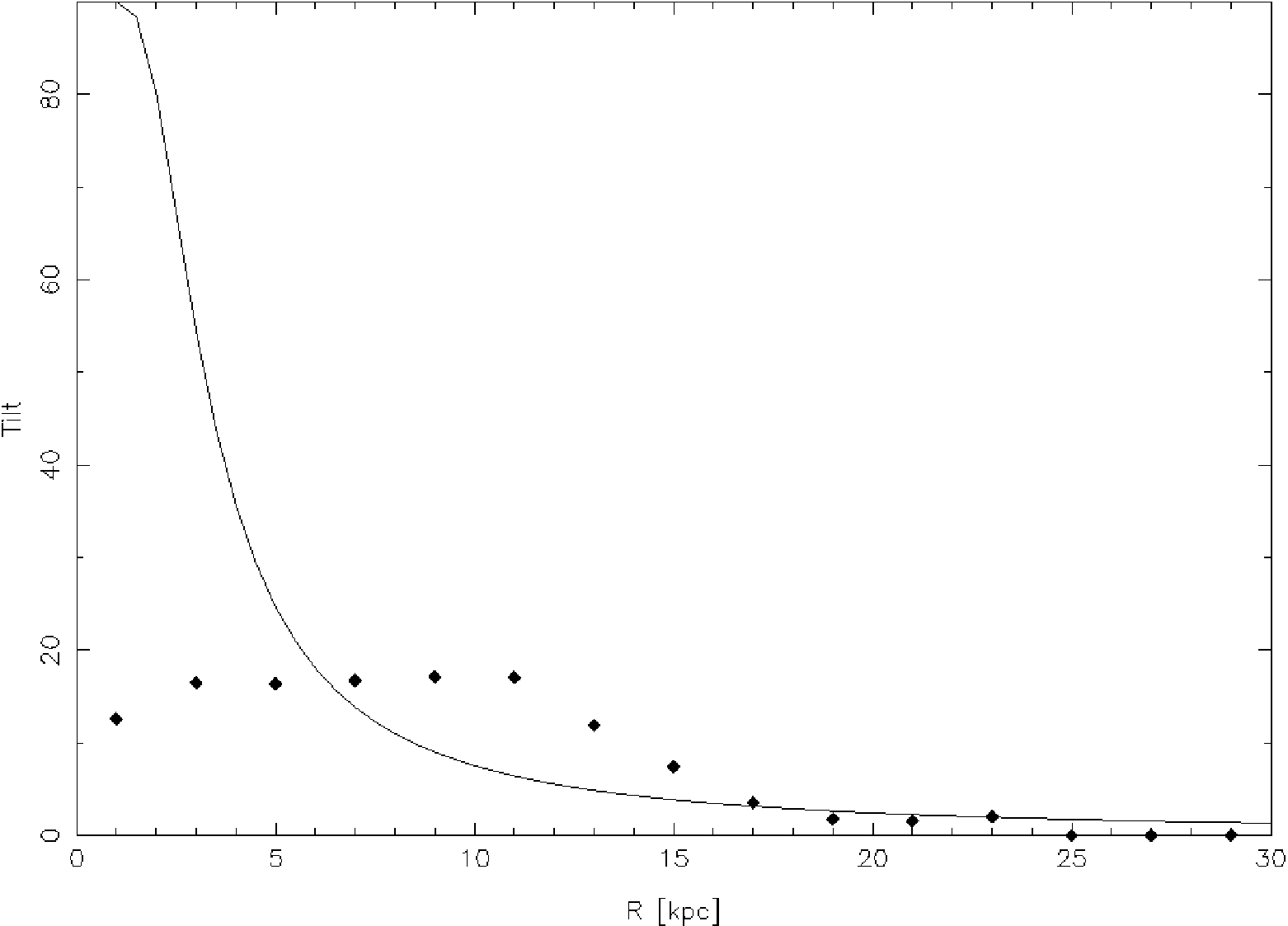,width=7.4cm,angle=0,clip=}}\par
}
\caption[]{a) \textit{(left)}
Simulation of a disk galaxy with mass $3\times 10^{10}\> M_\odot$
subjected to a typical cosmological torque for 1~Gyr. b) \textit{(right)}
Tilt of the disk from its original plane, computed in
spherical shells of width 2~kpc. The solid line is the expected behaviour
of a massless disk.}
\label{simulation disk}
\end{figure*}

Figure~\ref{simulation disk}a shows the simulation of a
 \( 3\times 10^{10}\: M_{\odot} \)
disk subjected to a torque for 1~Gyr. The disk was aligned with the $xy$-plane at
$t=0$. The inner part of the disk is flat and clearly tilted toward the positive \( x \)-axis.
Beyond 11~kpc, the disk warps back toward the original
plane. It distinctly resembles observed warped galaxies.
The particles of the simulated galaxy were binned into spherical shells 2~kpc
wide, and the minor axis of each bin was found from the moment of inertia tensor.
Figure~\ref{simulation disk}b plots the tilt angle of each ring from
the initial plane of the disk. The flat region shows up clearly as the inner
rings which are all tilted a uniform 17\jdegr\ from the initial plane,
while beyond 11~kpc (the {\sl warp radius}) the disk warps back
toward the initial plane.
A massless disk, in contrast,
does not exhibit an inner flat region and is warped at all radii, as shown by the solid line in
Figure~\ref{simulation disk}b. The self-gravity of the massive
disk maintains its flatness in its inner regions.

\begin{figure*}[tbp]
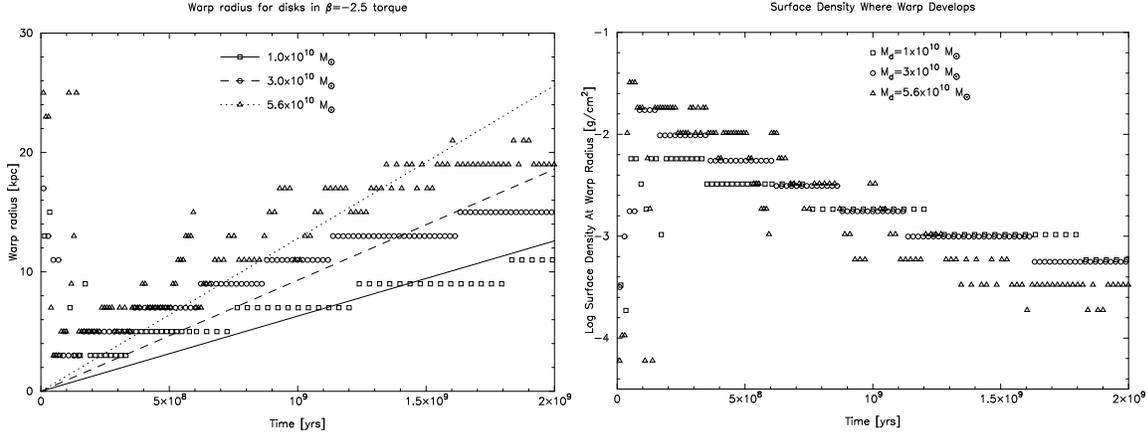

\vskip-2cm\hskip-5cm\vbox{\psfig{figure=bailf4.eps,height=7.5cm}\hskip2cm
  \psfig{figure=bailf5.eps,height=7.5cm}}\par
\caption[]{
a) \textit{(left)}
The symbols indicate the warp radius as a function of time for three disks of
mass $1\times 10^{10}\,$M$_\odot$, $3\times 10^{10}\,$M$_\odot$ and $5.6\times
10^{10}\,$M$_\odot$ respectively. 
Because the particles
were binned into spherical shells of width 2~kpc, the warp radii appear quantized.
The lines are linear least square fits. \label{rw2.5 figure}
b) \textit{(right)} For the same simulations as in (a), we show
the local surface density at the warp radius. As the warp moves out 
through the disk, the local surface
density at that radius falls. The local surface density at the warp radius is
similar for all models at each epoch.\label{plot surface density}
}
\end{figure*}

The warp radius $r_w$ at which the warp starts, 11~kpc in the disk shown in Figure~\ref{simulation disk},
is well-defined and can be followed over time.
It coincides with the radius at which the difference between the tilt of the
massive disk and the tilt of an equivalent massless disk is
maximized, i.e.~the data point that lies highest above the solid line
in Figure~\ref{simulation disk}b.
 As the disk evolves under the influence
of the torque, this warp radius moves out through the disk at a rate that depends
on the mass of the disk, as seen in Figure~\ref{rw2.5 figure}a. The three sets
of symbols represent simulations with disks of different mass.
Warps move faster through higher mass disks than through lower mass disks.

The effect of the disk mass on the rate of warp growth implies that the self-gravity
of the disk is important to the formation of the warp. In a massive disk,
particles at different radii are coupled to each other gravitationally and
act to keep the disk locally flat. This suggests that 
the crucial factor that determines
where the warp develops is the local surface density of the disk. In fact, 
Ostriker~\&
Binney~(1989) examined the effect of a slewing disk potential on a set of
self-gravitating rings and found that regions of high surface density react like
a solid body, but that warps can occur where the surface density is lower. Hofner~\&
Sparke~(1994) noted that in most cases the group speed of bending waves in
a disk of surface density $\Sigma(r)$ and angular rotation velocity $\Omega(r)$ is
$c_g = \frac{\pi G \Sigma(r)}{\Omega(r)}$, and therefore the time for a warp to
settle at a given radius is inversely proportional to the surface density at that
radius.

To test this hypothesis, we translate the warp radii $r_w$
of Figure~\ref{rw2.5 figure}a into local surface densities 
using $\Sigma (r_{w})=\frac{M_{d}}{4\pi r_{e}^{2}}e^{-r_{w}/r_{e}}$
for a disk of mass $M_d$ and exponential scale length $r_e$,
 and plot
the surface density at the warp radius as a function of time for the three
disks of different mass in Figure~\ref{plot surface density}b.
The surface density at the warp radius falls
as the warp moves out through the disk. Note that the local surface
densities at the warp \emph{at a given time} are quite similar for all models.
It appears that the local surface density is the important parameter for
determining the warp radius at a given time.

\section{Summary}
Our conclusions are the following:

\begin{itemize}
\item
Cosmological N-body simulations show that galactic disks are usually misaligned
with the mass distribution of the dark matter halo in which they are embedded.
Tidal torques are dominated by this misalignment 
and by asymmetries in the halo mass
distribution, such as satellite dwarfs. The resulting radial dependence of the
torque follows a power law of index -2.5. The gravitational torque is typically
strong enough to tilt the disk by 1~radian in 1--10~Gyr. Torques from mass
outside the halo appear unimportant.

\item
The tilting of the inner portions of the disk owing to the torque proceeds faster
than that of the outer regions, resulting in a warped disk. The self-gravity keeps the
inner regions of a massive disk  locally flat.

\item
The radius inside which the disk is flat grows with time. More massive disks have faster-growing
warps, mostly because only at large radius is the surface
density sufficiently low that its self-gravity cannot maintain the flat disk.
\end{itemize}

This work was supported by NSF grants AST98-07151 and PHY99-0749, and NSERC grant
PGSB-233028-2000.

{\small
\begin{description}{} \itemsep=0pt \parsep=0pt \parskip=0pt \labelsep=0pt
\item {\bf References}

\item
Binney, J. 1992, ARA\&A 30, 51
\item
Briggs, F. 1990, ApJ 352, 15
\item
Cole, S., Lacey, C. 1996, MNRAS 281, 716
\item
Debattista, V.P., Sellwood, J.A. 1999, ApJL 513, 107
\item
Dekel, A., Shlosman, I. 1983, in IAU Symposium 100, Internal Kinematics and Dynamics
	of Galaxies, ed. Athanassoula, E. Dordrecht: Kluwer, 187
\item
Hernquist, L. 1993, ApJS 86, 389
\item
Hofner, P., Sparke, L.S. 1994, ApJ 428, 466
\item
Navarro, J.F., Frenk, C.S., White, S.D.M.\ 1997, ApJ 490, 493
\item
Navarro, J.F., Steinmetz, M. 1997, ApJ 478, 13
\item
Ostriker, E.C., Binney, J.J. 1989, MNRAS 237, 785
\item
Quinn, T., Binney, J. 1992, MNRAS 255, 729
\item
Reshetnikov, V., Combes, F. 1998, A\&A 337, 9
\item
Sparke, L.S., Casertano, S. 1988, MNRAS 234, 873
\item
Steinmetz, M. 1996, MNRAS 278, 1005
\item
Toomre, A. 1983, in IAU Symposium 100, Internal Kinematics and Dynamics
	of Galaxies, ed. Athanassoula, E. Dordrecht: Kluwer, 177
\item
Warren, M.S., Quinn, P.J., Salmon, J.k., Zurek, W.H. 1992, ApJ 399, 405
\end{description}
}

\end{document}